\documentclass[a4paper,12pt,twoside]{article}

\usepackage[british]{babel}

\usepackage{amsmath,amssymb,amsfonts}
\usepackage{epsfig, color, subfig}
\usepackage{enumerate,booktabs,makecell}

\def\vet#1{{\boldsymbol #1}}
\def\build#1_#2^#3{\mathrel{\mathop{\kern 0pt#1}\limits_{#2}^{#3}}}

\def\interi{\mathbb{Z}}
\def\toro{\mathbb{T}}
\newcommand\ie{{\rm i.e.}, }
\newcommand\eg{, {\rm e.g.}, }

\def\Dscr{\mathcal{D}}

\def\Lbf{\mathbf{L}}

\def\Lscr{\mathcal{L}}
\def\Oscr{\mathcal{O}}
\def\Pscr{\mathcal{P}}

\def\Tscr{\mathcal{T}}

\def\epsilon{\varepsilon}
\def\rho{\varrho}
\def\phi{\varphi}

\def\poisson#1#2{\lbrace #1,#2 \rbrace}
\def\fastpoisson#1#2{\left\{ #1,#2 \right\}_{\vet{L},\vet{\lambda}}}
\def\secpoisson#1#2{\left\{ #1,#2 \right\}_{\vet{\xi},\vet{\eta}}}

\def\wideitem#1{\par\hangindent\itemindent
   \noindent\hbox to\parindent{\hfil{#1}\enspace}\ignorespaces}

\title{\bf A reverse KAM method to estimate\\ unknown mutual
  inclinations in\\ exoplanetary systems\thanks{\noindent{\it Key
      words and phrases:} exoplanets, n-body planetary problem, KAM
    theory, Celestial Mechanics.  {\it 2010 Mathematics Subject
      Classification.}  Primary: 70F15; Secondary: 70H08, 37N05,
    85A04, 85--08.}}

\author{
{\bf MARA VOLPI}\thanks{FRIA Research Fellow.}\\
{\small NaXys, Department of Mathematics, University of Namur}\\
{\small Rue Rempart de la Vierge 8, 5000\ ---\ Namur, Belgium.}\\
{\bf UGO LOCATELLI}\\
{\small Dipartimento di Matematica, 
Universit\`a degli Studi di Roma ``Tor Vergata'',}\\
{\small via della Ricerca Scientifica 1, 00133\ ---\ Roma, Italy.}\\
{\bf MARCO SANSOTTERA}\\
{\small Dipartimento di Matematica, Universit\`a degli Studi di Milano,}\\
{\small via Saldini 50, 20133\ ---\ Milano, Italy.}\\
{\small e-mails:
  {\tt mara.volpi@unamur.be, locatell@mat.uniroma2.it,}}\\
{\small {\tt marco.sansottera@unimi.it}}
}

\date{}

\begin{document}
\maketitle

\markboth{M. Volpi, U. Locatelli, M. Sansottera}{A reverse KAM method
  to estimate unknown mutual inclinations in exoplanetary
  ${\scriptscriptstyle\ldots}$}

\begin{abstract}
\noindent
The inclinations of exoplanets detected via radial velocity method are
essentially unknown.  We aim to provide estimations of the ranges of
mutual inclinations that are compatible with the long-term stability
of the system. Focusing on the skeleton of an extrasolar system, i.e.,
considering only the two most massive planets, we study the
Hamiltonian of the three-body problem after the reduction of the
angular momentum.  Such a Hamiltonian is expanded both in Poincar\'e
canonical variables and in the small parameter $D_2$, which represents
the normalised Angular Momentum Deficit. The value of the mutual
inclination is deduced from $D_2$ and, thanks to the use of interval
arithmetic, we are able to consider open sets of initial conditions
instead of single values.  Looking at the convergence radius of the
Kolmogorov normal form, we develop a \emph{reverse KAM approach} in
order to estimate the ranges of mutual inclinations that are
compatible with the long-term stability in a KAM sense. Our method is
successfully applied to the extrasolar systems HD~141399, HD~143761
and HD~40307.
\end{abstract}


\section{Introduction}
\label{sec:intro}

Nowadays, the number of catalogued multiple-planet systems is rapidly
reaching one thousand.  Their orbital characteristics are often quite
different with respect to those of our Solar System (for a review
containing a classification of the possible architectures see,
e.g.,~\cite{Winn-Fab-2015}).  In the present work, we will focus on
the analysis of the long-term evolution of the observed exoplanets, in
a spirit similar to the classical studies of stability of our Solar
System.

Multiplanetary extrasolar systems raise new interesting challenges
concerning the mathematical aspects of the orbital dynamics.  For
instance, in our Solar System the eccentricities of the celestial
bodies play the role of small parameters in the power series
expansions considered in classical perturbation theory. On the other
hand, the observed eccentricities of the major planets in extrasolar
systems are often so large (see, e.g.,~\cite{Butler-et-al-2006}), that
they prevent the convergence of the Laplacian expansion of the
disturbing function (see,
e.g.,~\cite{FerrazMello-1994}). Nevertheless, accurate analytical
results based on classical expansions have been obtained even for
systems having moderate eccentricities via high-order expansions (see,
e.g.,~\cite{Lib-San-2013}).

In the present work, we limit our study to the exoplanets observed via
radial velocity (hereafter, RV) method, because of its ability to
detect massive bodies. Therefore, RV based observations are expected
to capture information about the {\it skeleton} of an extrasolar
system, i.e., its major planets.  As a main drawback, the RV method
cannot detect the orbital inclinations; moreover, its measure of the
mass of each planet is affected by the uncertainty factor $\sin i$,
being $i$ the inclination of the plane of motion with respect to the
tangent plane to the celestial sphere (see,
e.g.,~\cite{Beau-FerM-Mich-2012}).  However, ranges of the most
probable values of the inclinations can be deduced by prescribing the
long-time stability of the system. This is done for instance
in~\cite{Las-Cor-2009}, where the properties of the numerically
computed orbital motions are investigated by using the frequency
analysis method (see~\cite{Laskar-03} and references therein for an
introduction to this kind of numerical explorations). We propose here
a novel procedure: a \emph{reverse KAM approach} by using
normal forms depending on a free parameter related to the unknown
mutual inclinations of the exoplanets. Our approach is based on a
careful adaptation of the algorithm constructing the Kolmogorov normal
form for the secular part of the Sun--Jupiter--Saturn (SJS) system
(see~\cite{Loc-Gio-2000}; see also~\cite{Kolmogorov-1954, Arnold-1963,
  Moser-1962}, that are the original articles giving the name to the
KAM~theorem). The differences between the two contexts are
remarkable. In~\cite{Loc-Gio-2000} the parameters and the orbital
elements of the SJS system were very well known; all these data were
used to prove the existence of KAM tori confining the motion and,
therefore, the stability of the secular model. Here, we deal with
systems for which some of the orbital elements are unknown: we aim to
infer information about their values by prescribing the stability and
therefore requiring that the algorithm constructing KAM tori is
convergent. Actually, from a practical point of view, its
implementation is rather delicate. For instance, we use the interval
arithmetic to represent the coefficients of the secular expansions;
this allows us to consider sets of values of the free parameter in a
comprehensive way instead of studying many different numerical
integrations, each corresponding to a single value of that same
parameter ranging in a suitably chosen discrete grid. Thus, our
implementation is an interesting example of alternative use of
validated numerics outside the context of a rigorous proof where it is
often used (see, e.g.,~\cite{Cel-Gio-Loc-2000}). We emphasise that
this is done by handling the difficulty due to the fact that the free
parameter, related to the unknown mutual inclination, directly
contributes to the so-called Laplace-Lagrange approximation (see,
e.g.,~\cite{Lib-San-2013}). Therefore, it affects the secular
frequencies possibly introducing dangerous resonance relations.

We believe that our approach can interestingly complement some recent
results: in particular, the concept of ``AMD-stability'' introduced
in~\cite{Las-Pet-2017} to analyse the dynamics of the multiple-planet
extrasolar systems (see also~\cite{Pet-Las-Bou-2017} for an extension
to the resonant case). Roughly, that criterion requires that the
Angular Momentum Deficit\footnote{The Angular Momentum Deficit is
  defined as the difference between the total value of the angular
  momentum and its value in the case of Keplerian circular coplanar
  orbits having radii equal to the semi-major axes of the planets.}
(hereafter AMD) is smaller than a critical threshold, in order to
ensure that the planetary orbits cannot collide; therefore, the system
is considered to be AMD-stable.  In~\cite{Las-Pet-2017} five planetary
systems are recognised to belong to the so-called subcategory of
``{\it hierarchical AMD-stable systems that are AMD-unstable but
  become AMD-stable when they are split into two parts}''. Among them,
our Solar System is a typical example when considering the two
subsystems formed by the giant planets on one side and the inner ones
on the other. We emphasise that AMD-stability of the giant planets is
not sufficient to prove the global stability of the system as it does
not provide a detailed enough information about the regularity of
their motions. Indeed, it is well known that the chaoticity of the
secular motions of the inner planets is induced by the gravitational
perturbations due to Jupiter (see~\cite{Laskar-1990}). Because of this
chaoticity, it has been possible to select some scenarios (depicted by
suitably chosen numerical integrations) leading to the ejection of
Mercury or to destructive collisions between the terrestrial planets
in a few billions of years (see~\cite{Laskar-1989c} in the context of
the secular dynamics and~\cite{Las-Gas-2009}, respectively).  It is
very natural to expect that these destabilising effects would act
dramatically faster, if also the secular dynamics of the outer system
were chaotic, instead of being extremely regular as it has shown to be
(see, e.g.,~\cite{Laskar-1996} also as a review covering most of the
properties of the Solar System that have been briefly recalled here).
A deeper knowledge of the dynamics of the outer planets is therefore
crucial in order to prove the effective (or long-time) stability of
the complete system. When a specific extrasolar system cannot be
classified as globally AMD-stable, the problem of ensuring its
stability properties can be attacked by following a strategy that is
based on our reverse KAM approach, as outlined below.

In the case of hierarchical AMD-stable systems, when successful our
approach can ensure that there are values of the inclinations for
which the subsystem formed by the major planets is stable in a much
stronger sense with respect to the AMD-criterion: the eventual
diffusion would be so weak that the orbit could not significantly go
away from a KAM torus before an extremely\footnote{ Actually, when the
  mild hypotheses assumed in~\cite{Mor-Gio-1995} are satisfied, the
  diffusion time is estimated to be super-exponentially big. This
  means that its order of magnitude is given by the exponential of the
  exponential of the inverse of a fractional power of the distance
  from a reference KAM torus.} long interval of time
(see~\cite{Gio-Loc-San-2017}). In such a situation, the motion of the
biggest planets is indistinguishable from a quasi-periodic one. Such a
preliminary result would be essential in order to prove (at least) the
metastability of the less massive planets over times that are
comparable with the expected lifetime of the system. This highlights
the usefulness of our reverse KAM approach.

In the present paper, we apply our KAM-stability to three extrasolar
systems that are modeled by a three-body approximation, which includes
the star and the two biggest planets. Quite remarkably, one of these
systems, HD~141399, is hierarchical AMD-stable according to the
classifications given in~\cite{Las-Pet-2017}; another one, HD~40307,
is included in the category of the AMD-unstable systems. This work
represents the first step in the direction of a complete proof of the
so-called \emph{effective stability} of such exoplanetary systems,
when they are studied in the framework of models including all their
already discovered planets (see~\cite{San-Loc-Gio-2013} for a recent
application of these concepts to the secular dynamics of the
Sun--Jupiter--Saturn--Uranus system). Of course, the whole
implementation of our strategy is not priceless: the required amount
of computations (mainly by the algebraic manipulations of the
expansions) is extremely demanding.

Our paper is organised as follows.  In Sect.~\ref{sec:model} we recall
the initial expansions of the secular Hamiltonian model of the
three-body planetary problem. In Sect.~\ref{sec:invtori} we deal with
the algorithm constructing invariant tori for such a model. In
Sect.~\ref{sec:compalg} we describe the way to infer information about
the range of values of the mutual inclination. The applications of our
method to three extrasolar systems are discussed in
Sect.~\ref{sec:res}. Finally, the conclusions are outlined in
Sect.~\ref{sec:conc}.

\section{Settings for the definition of the Hamiltonian model}
\label{sec:model}

As it has been mentioned in the Introduction our approach is based on
a careful adaptation of that described
in~\cite{Loc-Gio-2000}. However, for the sake of completeness, it is
convenient to briefly recall both the definitions and the preliminary
canonical transformations that properly introduce the secular model we
are going to study.

\subsection{The expansion of the Hamiltonian}
\label{subsec:exp}
Object of this study is a three-body planetary problem, formed by a
central star (indicated by the index $0$) and two planets revolving
around it, marked by the indexes $1$ (inner) and $2$ (outer).  Thanks
to the conservation of the angular momentum, we can perform the
reduction of the nodes which allows us to use the planar Poincar\'e
variables
\begin{equation}
  \begin{aligned}
    \Lambda_j &= \frac{m_0 m_j}{m_0+m_j} \sqrt{(m_0 + m_j) a_j}\qquad & \lambda_j &=M_j + \omega_j\\
    \xi_j &=\sqrt{2\Lambda_j} \sqrt{1-\sqrt{1-e_j^2}} \cos(\omega_j)
    &\quad \eta_j &= -\sqrt{2\Lambda_j} \sqrt{1-\sqrt{1-e_j^2}} \sin(\omega_j)\\
  \end{aligned}
\end{equation}
where $a_j$, $e_j$, $M_j$ and $\omega_j$ are the semi-major axis, the
eccentricity, the mean anomaly and perihelion argument of the $j$-th
planet, respectively.  We introduce the translation $L_j=\Lambda_j -
\Lambda^*_j$, where $\Lambda^*_j$ is the value of $\Lambda_j$ for the
       {\it observed} semi-major axis $a_j$.

Following~\cite{Robutel-1995}, we expand the Hamiltonian both in the
Poincar\'e variables and in the parameter $D_2\,$, that is a sort of
``normalised angular momentum deficit'',
\begin{equation}
  D_2=\frac{(\Lambda_1+\Lambda_2)^2-C^2}{\Lambda_1\Lambda_2}
\end{equation}
where $C$ is the norm of the total angular momentum: by definition,
$D_2$ is ${\cal O}(e^2+i^2)$\,. This parameter is therefore a measure
of the difference between the actual total angular momentum and the
one of a similar system having circular and co-planar orbits (for
which $D_2 = 0$). As a main difference with respect to the approach
in~\cite{Loc-Gio-2000}, which we constantly refer to, here we keep
$D_2$ as a free parameter in the expansions, while there it was
replaced by its particular value (computed for the
Sun-Jupiter-Saturn system).  The Hamiltonian expansion in power series
of the variables $\Lbf$, $\boldsymbol{\xi}$, $\boldsymbol{\eta}$,
parameter $D_2$ and in Fourier series of $\boldsymbol{\lambda}$ writes
\begin{equation}
  \label{eq:ham0}
  H^{(\Tscr_F)}=\sum_{j_1=1}^{\infty}h^{({\rm Kep})}_{j_1,0}(\Lbf)
  + \mu\sum_{s=0}^{\infty}\sum_{j_1=0}^{\infty}\sum_{j_2=0}^{\infty}D_2^s\, h^{(\mathcal{T}_F)}_{s;j_1,j_2}(\Lbf,\boldsymbol{\lambda},\boldsymbol{\xi},\boldsymbol{\eta})
\end{equation}
where $\mu = \max\{m_1/m_0,m_2/m_0\}$ and
\begin{itemize}
\item $h^{({\rm Kep})}_{j_1,0}$ is a homogeneous polynomial function
  (hereafter h.p.f.) of degree $j_1$ in $\Lbf$; in particular,
  $h^{({\rm Kep})}_{1,0} = \vet{n}^*\cdot\Lbf$, where the components
  of the angular velocity vector $\vet{n}^*$ are defined by the third
  Kepler law.
\item$h^{(\Tscr_F)}_{s;j_1,j_2}$ is a h.p.f. of degree $j_1$ in
  $\Lbf$, degree $j_2$ in $\boldsymbol{\xi}$ and $\boldsymbol{\eta}$,
  and with coefficients that are trigonometric polynomials in
  $\boldsymbol{\lambda}$.
\end{itemize}

\subsection{The secular Hamiltonian at order two in the masses}
\label{subsec:order2}
The main idea of this work is based on the construction of invariant
tori through the application of a Kolmogorov normalisation
algorithm. The Kolmogorov normalisation scheme is also adapted to
preliminary produce the secular approximation at order two in the
masses (see, e.g.,~\cite{Loc-Gio-2000, Lib-San-2013}).  This means
that in our model the torus corresponding to $\Lbf = 0$ in the new
coordinates will be invariant up to order two in the masses. For this
aim, we proceed by averaging over the fast angles the terms of the
Hamiltonian~\eqref{eq:ham0} that do not depend or are linear in the
actions $\Lbf$. This elimination is obtained via a composition of two
Kolmogorov-like steps.

First, the transformed Hamiltonian writes, in the Lie series formalism,
\begin{equation}
  \exp\Lscr_{\chi^{(\Oscr 2)}_1}H^{(\Tscr_F)}=\sum_{j=0}^{\infty}\frac{1}{j!}\Lscr^{j}_{\chi^{(\Oscr 2)}_1}H^{(\Tscr_F)}\,,
\end{equation}
where the generating function $\chi^{(\Oscr 2)}_1$ is determined as
the solution of the following homological equation
\begin{equation}
  \label{eq:homeq}
  \sum_{i=1}^2n^*_i\cdot\frac{\partial \chi^{(\Oscr 2)}_1}{\partial
    \lambda_i}+ \mu \sum_{{s=0\,,\>j_2=0}\atop{2s+j_2\le
      N_S}}\left\lceil D_2^s\, h_{s;0,j_2}^{(\Tscr_F)}\right\rceil_%
      {\boldsymbol{\lambda}:K_F}= \mu
      \sum_{{s=0\,,\>j_2=0}\atop{2s+j_2\le N_S}}D_2^s\, \Big\langle
      h_{s;0,j_2}^{(\Tscr_F)}\Big\rangle_{\boldsymbol{\lambda}}\,,
\end{equation}
being $\langle \cdot \rangle_{\vet{\phi}}$ the average over the
generic angles $\vet{\phi}$.  In the previous formula, we have denoted
with $\lceil g \rceil_{\boldsymbol{\lambda}:K_F}$ the truncation of
the expansion of the generic function $g$ up to a trigonometric degree
$K_F\,$. The parameter $K_F$ is fixed so as to include the main
quasi-resonance of the system on hand: for instance, let us suppose
the system is near to a $k_1^*:k_2^*$ resonance, then we set $K_F \ge
|k_1^*|+|k_2^*|\,$.  Moreover, in~\eqref{eq:homeq} the integer
parameter $N_S$ rules the considered order of magnitude in
eccentricity and inclination: the choice of the particular value of
$N_S$ is again related to the main quasi-resonance of the system. In
fact, for the D'Alembert rules we know that the terms containing
harmonics $(k_1^*\lambda_1 - k_2^*\lambda_2)$ have order in
eccentricity and inclination greater or equal than $|k_1^*|-|k_2^*|$
and with the same parity. Therefore, in order to include the effects
of the $k_1^*:k_2^*$ resonance in the generating function
$\chi^{(\Oscr 2)}_1$, we have to truncate the expansion up to $N_S \ge
|k_1^*|-|k_2^*|$. This constraint takes into account that both
$\boldsymbol{\xi}$ and $\boldsymbol{\eta}$ are ${\cal O}(e)$ and $D_2$
is ${\cal O}(e^2+i^2)$.  To fix the ideas, let us focus on the main
extrasolar planets ${\rm HD~141399~c}$ and ${\rm HD~141399~d}$, whose
periods are approximately equal to $202$~days and $1070$~days,
respectively. Therefore, the quasi-resonance $5:1$ is expected to
substantially affect the dynamics: according to the previous
discussion, we then fix $K_F \ge 6$ and $N_S \ge 4$. Of course, these
criteria determine just the {\it lower} bounds on the integer
parameters $K_F$ and $N_s\,$: we stress that one could be interested
in producing larger expansions according to the available
computational power.  In Section \ref{sec:res}, Tab.~\ref{tab:parexp}
we will list the particular value of $K_F$ and $N_S$ for each of the
systems considered by our applications.

The second Kolmogorov-like step is performed in an analogous way so as to
introduce the normalised Hamiltonian up to order two in the masses
$H^{(\Oscr 2)}=\exp\Lscr_{\chi^{(\Oscr 2)}_2}\,\circ\,\exp\Lscr_{\chi^{(\Oscr 2)}_1}H^{(\Tscr_F)}\,$,
where the new generating function $\chi^{(\Oscr 2)}_2$ is the solution
of the  homological equation
\begin{equation}
  \sum_{i=1}^2n^*_i\cdot\frac{\partial \chi^{(\Oscr 2)}_2}{\partial\lambda_i}+
  \mu \sum_{{s=0\,,\>j_2=0}\atop{2s+j_2\le N_S}}\left\lceil D_2^s\,h_{s;1,j_2}^{(\Tscr_F)}\right\rceil_%
  {\boldsymbol{\lambda}:K_F} + {\Lscr_{\chi^{(\Oscr 2)}_1}h^{({\rm Kep})}_{2,0}} = \mu \sum_{{s=0\,,\>j_2=0}\atop{2s+j_2\le N_S}}D_2^s\,\Big\langle h_{s;1,j_2}^{(\Tscr_F)}\Big\rangle_{\boldsymbol{\lambda}}.
\end{equation}

As we already mentioned, we will focus on the secular part of the
Hamiltonian $\langle H^{(\Oscr 2)} \rangle_{\vet{\lambda}}$: for such
an Hamiltonian, the actions $\Lbf$ are first integrals. We consider
the basic approximation of the fast dynamics corresponding to
quasi-periodic motions with an angular velocity vector equal to
$\vet{n}^*$, by setting $\Lbf = 0$.

Let us define
\begin{equation}
  \label{eq:htilde}
  \begin{aligned}
    \widetilde{\kern-2pt H} &= H^{(\Tscr_F)} +
    \frac{1}{2}\fastpoisson{\chi^{(\Oscr 2)}_1}{\Lscr_{\chi^{(\Oscr 2)}_1}h^{({\rm Kep})}_{2,0}}\\
    &+\fastpoisson{\chi^{(\Oscr 2)}_1}{\mu \sum_{{s=0\,,\>j_2=0}\atop{2s+j_2\le N_S}} D_2^s\,\tilde{h}_{s;1,j_2}^{(\Tscr_F)}}+
    \frac{1}{2}\secpoisson{\chi^{(\Oscr 2)}_1}{\mu \sum_{{s=0\,,\>j_2=0}\atop{2s+j_2\le N_S}}D_2^s\,\tilde{h}_{s;0,j_2}^{(\Tscr_F)}}
  \end{aligned}
\end{equation}
where $\fastpoisson{\cdot}{\cdot}$ and $\secpoisson{\cdot}{\cdot}$ are
the terms of the Poisson bracket involving only the derivatives with
respect to the variables $(L,\lambda)$ and $(\xi,\eta)$,
respectively. Then, according to~\cite{Loc-Gio-2000}, we have that
\begin{equation*}
  \langle H^{(\Oscr 2)} \rangle_{\vet{\lambda}} \Big|_{\Lbf= \vet{0}} =
\langle\, \widetilde{\kern-2pt H}\, \rangle_{\vet{\lambda}} \Big|_{\Lbf= \vet{0}} + {\cal O} (\mu^3)\,.
\end{equation*}
We can finally introduce our secular model up to order two in the
masses, by setting
\begin{equation}
  \label{eq:hamsec}
  H^{({\rm sec})}(D_2,\boldsymbol{\xi},\boldsymbol{\eta}) =
  \left\lceil\,
  \langle\,\widetilde{\kern-2pt H}\,\rangle_{\vet{\lambda}}
  \Big|_{\Lbf= \vet{0}}\,\right\rceil_{(D_2,\boldsymbol{\xi},\boldsymbol{\eta})\,:\,2N_S}\, ,
\end{equation}
where $\big\lceil\,\langle\,\widetilde{\kern-2pt
  H}\,\rangle_{\vet{\lambda}} \big|_{\Lbf=
  \vet{0}}\,\big\rceil_{(D_2,\boldsymbol{\xi},\boldsymbol{\eta})\,:\,2N_S}$
indicates the averaged expansion (over the fast angles
  $\vet{\lambda}$) of the part of $\widetilde{\kern-2pt H}$ that is
  both independent from the actions $\Lbf$ and truncated up to a total
  order of magnitude $N_S$ in eccentricity and inclination. This
means that a monomial $D_2^s\, \vet{\xi}^{\vet{m}_1}
\vet{\eta}^{\vet{m}_2}$ is included in the truncation if and only if
$2s+|\vet{m}_1|+|\vet{m}_2| \le 2N_S\,$.  By comparing
\eqref{eq:htilde} and \eqref{eq:hamsec}, one can notice that our
secular Hamiltonian model represented by $H^{({\rm sec})}$ does not
depend on the second generating function $\chi^{(\Oscr 2)}_2$ whose
explicit calculation is therefore unnecessary.

The explicit form of \eqref{eq:hamsec} writes
\begin{equation}
  \label{eq:hamsecfin}
  H^{({\rm sec})} =
  h_{1,1}^{({\rm sec})}\,+\sum_{s=2}^{N_S}\sum_{l=1}^{s}D_2^{s-l} h_{s,l}^{({\rm sec})}
\end{equation}
where $h_{s,l}$ is a homogeneous polynomial function of degree $2l$ in
$\boldsymbol{\xi}$ and $\boldsymbol{\eta}\,$, $\forall\ 1\le l\le s
\le N_S\,$.  The even parity of the exponents is determined by the
D'Alembert rules: having removed all the harmonics, the order in
eccentricity that the terms must held is of the same parity of zero.
The expansion of the final Hamiltonian $H^{({\rm sec})}$ presents
terms in $D_2$, $\boldsymbol{\xi}$ and $\boldsymbol{\eta}$ up to a
degree that is twice the one of the truncated expansions of
$\chi^{(\Oscr 2)}_1$ as it is determined by \eqref{eq:homeq}: this is
set to ensure that all the terms generated by the Poisson brackets in
\eqref{eq:htilde} are going to be taken into account.

\section{Construction of invariant tori for our secular model}
\label{sec:invtori}

\subsection{Preliminary set up for the Kolmogorov algorithm}
We will perform a series of preliminary transformations in order to
obtain the most convenient formulation of our Hamiltonian for the
construction of the invariant torus.  Firstly, we will diagonalise the
quadratic part of the Hamiltonian; secondly, we will transform the
variables into an action-angle set; we will then proceed with a
partial Birkhoff's normalisation, so as to remove the degeneration of the
unperturbed Hamiltonian; finally, we will shift the origin of the
actions so that they are centred around a value consistent with the
observations.

It is well known that under mild assumptions on the quadratic part of
the Hamiltonian which are satisfied in our case (see Sect.~3
of~\cite{Bia-Chi-Val-2006} where such hypotheses are shown to be
generically fulfilled for a planar model of our Solar System) one can
find a canonical transformation
$(\vet{\xi},\vet{\eta})=\Dscr(\vet{x},\vet{y})$ with the following
properties: (i)~the map
$(\vet{\xi},\vet{\eta})=~\big(\vet{\xi}(\vet{x}),\vet{\eta}(\vet{y})\big)$
is linear, (ii)~$\Dscr$ diagonalizes the quadratic part of the
Hamiltonian, so that we can write $h_{1,1}^{({\rm sec})}$ in the new
coordinates as $\sum_{j=1}^2\nu_j(x_j^2+y_j^2)/2\,$, where both the
entries of the vector $\vet{\nu}$ have the same sign.

Action--angle variables are introduced via the
canonical transformation
\begin{equation}
x_j=\sqrt{2I_j}\cos \phi_j\ ,
\qquad
y_j=\sqrt{2I_j}\sin \phi_j\ ,
\qquad j=1,2\ .
\label{eq:azang}
\end{equation}
With these two last changes of coordinates the
Hamiltonian~\eqref{eq:hamsecfin} takes the form
\begin{equation}
  H^{({\rm I})}(\boldsymbol{I},\boldsymbol{\phi})=  \boldsymbol{\nu}\cdot\boldsymbol{I} \,+\,\sum_{s=2}^\infty\sum_{l=1}^{s}
  D_2^{s-l}h_{s;l}^{({\rm I})}(\boldsymbol{I},\boldsymbol{\phi}) \ ,
\label{eq:hamsecazang}
\end{equation}
where $h_{s;l}$ is an homogeneous polynomial function of degree $2l$
in the square roots of actions $\boldsymbol{I}$ and a trigonometric
polynomial of degree $2s$ in angles $\boldsymbol{\phi}\,$, \ie it
writes
\begin{equation}
  h_{s,l}^{({\rm I})}(\vet{I},\vet{\phi})=
  \sum_{i_1+i_2=2l}\,\sum_{j_1=0}^{i_1}\sum_{j_2=0}^{i_2}
  c_{s;i_1;i_2;j_1;j_2}^{({\rm I})}
  \sqrt{I_1^{i_1}I_2^{i_2}}\cos\big[(i_1-2j_1)\phi_1+(i_2-2j_2)\phi_2\big]\,.
\end{equation}
In the previous formula only cosines occur because of the parity
relation due to the D'Alembert rules.

Let us stress that our aim is to provide ranges of inclinations which
are compatible with the stability of the system.  These intervals of
values are obtained as a function of the angular momentum deficit
parameter $D_2$.  Thus it is crucial to keep $D_2$ as a parameter in
the Hamiltonian expansion as long as possible.  We now proceed with a
partial Birkhoff's normalisation in order to remove the degeneration
of the unperturbed Hamiltonian. We can visualise the Hamiltonian
\eqref{eq:hamsecazang} as
\begin{equation}
  \label{eq:hamscheme}
  \vcenter{\openup1\jot\halign{
      \hbox to 14 ex{\hfil $\displaystyle {#}$\hfil}
      &\hbox to 10 ex{\hfil $\displaystyle {#}$\hfil}
      &\hbox to 10 ex{\hfil $\displaystyle {#}$\hfil}
      &\hbox to 10 ex{\hfil $\displaystyle {#}$\hfil}
      &\hbox to 10 ex{\hfil $\displaystyle {#}$\hfil}
      &\hbox to 10 ex{\hfil $\displaystyle {#}$\hfil}
      &\hbox to 4 ex{\hfil $\displaystyle {#}$\hfil}
      \cr
      & & & & & {\scriptstyle \build{\phantom{\cdot}}_{\cdot}^{}
                  \phantom{\cdot}\cdot\phantom{\cdot}
                  \build{\phantom{\cdot}}_{}^{\cdot} } 
      \cr
      & & & & h_{4;4}^{({\rm I})} & \ldots
      \cr
      & & & h_{3;3}^{({\rm I})} & D_2\,h_{4;3}^{({\rm I})} & \ldots
      \cr
      & & h_{2;2}^{({\rm I})} & D_2\,h_{3;2}^{({\rm I})} & D_2^2\,h_{4;2}^{({\rm I})} & \ldots
      \cr
      H^{({\rm I})}(\boldsymbol{I},\boldsymbol{\phi})= & \vet{\nu}\cdot\vet{I} & D_2\,h_{2;1}^{({\rm I})}
      & D_2^2\,h_{3;1}^{({\rm I})} & D_2^3\,h_{4;1}^{({\rm I})} & \ldots &.
      \cr
  }}
\end{equation}
This writing highlights two features of each term: the size of the
perturbation in eccentricity and inclination is determined by the
columns; the degree in actions depends on the rows.  Our aim is then
to remove the dependency on the angle variables up to the third
column.  We determine the first generating function by solving
\begin{equation}
  \label{eq:genfun1}
\poisson{{\cal B}_1^{({\rm II})}}{\,\vet{\nu}\cdot\vet{I}} - D_2\,h_{2;1}^{({\rm I})} = D_2 Z_{2;1}\,
\end{equation}
where $Z_{s,l}$ is the average of $h_{s;l}^{({\rm I})}$ over the angles $\vet{\phi}$.

In the same way, we determine the generating function ${\cal
  B}_2^{({\rm II})}$ as the solution of
\begin{equation}
  \label{eq:genfun2}
  \poisson{{\cal B}_2^{({\rm II})}}{\,\vet{\nu}\cdot\vet{I}} - h_{2;2}^{({\rm I})} = Z_{2;2}\,.
\end{equation}

The transformed Hamiltonian is computed as
\begin{equation}
  H^{({\rm II})}=\exp\Lscr_{{\cal B}^{({\rm II})}_2}\circ\exp\Lscr_{{\cal B}^{({\rm II})}_1}H^{({\rm I})}\,.
\end{equation}
Let us stress that $\exp\Lscr_{{\cal B}^{({\rm II})}_1}H^{({\rm I})}$
does not produce any contribution to the term $h_{2;2}^{({\rm I})}$:
this justifies the term appearing in~\eqref{eq:genfun2} for the
generating function.  At this point, all the terms up to order~4 in
eccentricity and inclination do not depend on the fast angles and the
Hamiltonian reads
\begin{equation}
  H^{({\rm II})}(\boldsymbol{I},\boldsymbol{\phi})=  \boldsymbol{\nu}\cdot\boldsymbol{I} \,+\,
  D_2\,Z_{2;1}(\vet{I}) + Z_{2;2}(\vet{I})+
  \sum_{s=3}^\infty\sum_{l=1}^{s}
  D_2^{s-l}h_{s;l}^{({\rm II})}(\boldsymbol{I},\boldsymbol{\phi}) \,.
\end{equation}

Analogously, we compute the generating functions ${\cal B}^{({\rm
    III})}_1$, ${\cal B}^{({\rm III})}_2$, ${\cal B}^{({\rm III})}_3$
in order to eliminate the dependency on the angle variables of the
terms of order~6 in eccentricity and inclinations. Finally, our
Hamiltonian is computed as
\begin{equation}
  H^{({\rm III})}=\exp\Lscr_{{\cal B}^{({\rm III})}_3}\circ\exp\Lscr_{{\cal B}^{({\rm III})}_2}\circ
  \exp\Lscr_{{\cal B}^{({\rm III})}_1}H^{({\rm II})}\,.
\end{equation}

The last preliminary transformation of the Hamiltonian consists in a
translation of the actions. Being the action vector $\vet{I}$ nearly
constant, i.e., $\vet{I}(t)\simeq\vet{I}(0)$, we shift the origin of
the action about $\vet{I}(0)=\vet{I}^*$.  This is done using a
canonical transformation $\Tscr(\boldsymbol{I},\boldsymbol{\phi})=
(\boldsymbol{p}+\boldsymbol{I}^*,\boldsymbol{q})$. The transformed
Hamiltonian is given by
\begin{equation*}
H^{(0)}(\vet{p},\vet{q})=H^{({\rm III})}\circ\Tscr(\vet{I},\vet{\phi})\,.
\end{equation*}
Let us remark that in~\cite{Loc-Gio-2000} the translation was
determined in such a way to construct a torus with a specific
frequency $\vet{\omega}$; that frequency was accurately computed from
the numerical integration via Fourier analysis.  The same approach
cannot be adopted in the present work: being the knowledge of the
parameters of the system only partial any numerically integration is
unattainable.

\subsection{Formal construction of the Kolmogorov invariant tori}
\label{subsec:tori}
We will now proceed with the construction of the Kolmogorov invariant
tori.  Firstly, we expand the Hamiltonian $H^{(0)}$,
whose expansion can be visually arranged as
\begin{equation}
\label{eq:hamexp}
  \vcenter{
    \halign to\hsize{
      &\hfil$\displaystyle\>{#}$\hfil
      &\ \ \hfil$\displaystyle\>{#}$\hfil
      &\ \ \hfil$\displaystyle\>{#}$\hfil
      &\ \ \hfil$\displaystyle\>{#}$\hfil
      &\ \ \hfil$\displaystyle\>{#}$\hfil
      &\ \ \hfil$\displaystyle\>{#}$\hfil
      &\ \ \hfil$\displaystyle\>{#}$\hfil
      \cr
      & &\vdots &\vdots&\vdots&\vdots&\vdots
      \cr
      & &f_2^{(0,0)(\boldsymbol{p})}
      &f_2^{(0,1)}(\boldsymbol{p},\boldsymbol{q})
      &\ldots
      &f_2^{(0,s)}(\boldsymbol{p},\boldsymbol{q})
      &\ldots
      \cr
      &H^{(0)}(\boldsymbol{p},\boldsymbol{q})\,=\,
      \Bigg.\sum &\boldsymbol{\omega}^{(0)}\cdot\boldsymbol{p}
      &f_1^{(0,1)}(\boldsymbol{p},\boldsymbol{q})
      &\ldots
      &f_1^{(0,s)}(\boldsymbol{p},\boldsymbol{q})
      &\ldots &\qquad ,
      \cr
      & &0\,
      &f_0^{(0,1)}(\boldsymbol{q})\,
      &\ldots
      &f_0^{(0,s)}(\boldsymbol{q})\,
      &\ldots
      \cr
    }
  }
\end{equation}
being the generic term $f_j^{(0,s)}\in\Pscr_{j,2s}$\,. This means that
it is a homogeneous polynomial of degree $j$ in the actions
$\boldsymbol{p}$ and a trigonometric polynomial of degree $2s$ in
$\boldsymbol{q}$. Therefore it is possible to represent such type of
terms on a computer because it is finite.  There is a striking
difference between the visual schemes~\eqref{eq:hamscheme}
and~\eqref{eq:hamexp}: in the latter, we do not keep track of the
expansions in powers of~$D_2$\,. This is due to the fact that, in the
explicit applications, we replace the parameter $D_2$ with convenient
intervals of values. In Sec.~\ref{sec:compalg} we will discuss in more
detail this technical point, that is not essential for the comprehension
of the normalisation scheme.

Let us emphasise that the terms $f_j^{(0,s)}$ in the $s$-th column are
of order $\|\vet{I^*}\|^s$, as it is discussed\eg
in~\cite{Gio-Loc-San-2017}. Therefore the parameter $\vet{I^*}$ rules
the convergence of the series with respect to the index $s$; according
to the definitions in the previous sections, it is a small quantity
because $\vet{I^*}$ is $\Oscr(e^2+i^2)\,$.

The Kolmogorov's normalisation algorithm requires to remove all the terms
of the Hamiltonian~\eqref{eq:hamexp} of degree~$0$ or~$1$ in the actions  
$\boldsymbol{p}$, with the exception of the term
$\boldsymbol{\omega}\cdot\boldsymbol{p}\,$. In order to do that, we start by
determining the generating function $\chi_1^{(1)}$ such that
\begin{equation}
\label{eq:kolgen1}
  \poisson{\chi_1^{(1)}}{\boldsymbol{\omega}^{(0)}\cdot\boldsymbol{p}}+
  f_0^{(0,1)}=0\,,
\end{equation}
where $\chi_1^{(1)}$ is a trigonometric polynomial of degree~$2$.

We will then obtain a new Hamiltonian
\begin{equation}
\hat H^{(1)}=\exp\Lscr_{\chi_1^{(1)}}H^{(0)}
\end{equation}
whose generic term of the expansion is $\hat
f_j^{(1,s)}\in\Pscr_{j,2s}$\,. As a consequence of
equation~\eqref{eq:kolgen1}, we have that $\hat f_0^{(1,1)}=0$\,.
 
We proceed in an analogous way to complete this first Kolmogorov's
normalisation step: we compute the generating function
$\chi_2^{(1)}(\boldsymbol{p},\boldsymbol{q})$ such that
\begin{equation}
\label{eq:kolgen2}
  \poisson{\chi_2^{(1)}}{\boldsymbol{\omega}^{(0)}\cdot\boldsymbol{p}}+
  \hat f_1^{(1,1)}=\langle\hat f_1^{(1,1)}\rangle_{\boldsymbol{q}}\,;
\end{equation}
then, $\chi_2^{(1)}$ will be linear in $\boldsymbol{p}$ and of order~2
in $\boldsymbol{q}$. Let us stress that it is possible to solve the
previous homological equations~\eqref{eq:kolgen1}
and~\eqref{eq:kolgen2}, provided that $|\vet{k} \cdot
\vet{\omega}^{(0)}| > 0$ for $\vet{k}\in\interi_2$ with
$|\vet{k}|=1,2$\,, being $|\vet{k}| = |k_1| + |k_2|$\,.

Therefore, we will obtain the new Hamiltonian
$H^{(1)}=\exp\Lscr_{\chi_2^{(1)}}\hat H^{(1)}$, whose generic term is
now $f_j^{(1,s)}\in\Pscr_{j,2s}\,$.  In the following it lies a
profound difference with respect to previous works (see for
example~\cite{Loc-Gio-2000}): due to the way $\chi_2^{(1)}$ was
determined, we have that
\begin{equation}
  f_1^{(1,1)}= \hat f_1^{(1,1)}+\Lscr_{\chi_2^{(1)}}%
  \boldsymbol{\omega}^{(0)}\cdot\boldsymbol{p} =%
  \langle\hat f_1^{(1,1)}\rangle_{\boldsymbol{q}}\,.
\end{equation}
Therefore, $f_1^{(1,1)}$ is an homogeneous polynomial of degree~$1$ in
$\boldsymbol{p}$ and independent from $\boldsymbol{q}$: hence, it
shares the same functional properties of the term
$\boldsymbol{\omega}^{(0)}\cdot\boldsymbol{p}\,$.  We then set for
appropriate values of $\boldsymbol{\omega}^{(1)}$
\begin{equation}
\boldsymbol{\omega}^{(1)}\cdot\boldsymbol{p}=%
\boldsymbol{\omega}^{(0)}\cdot\boldsymbol{p} +%
 \langle\hat f_1^{(1,1)}\rangle_{\boldsymbol{q}}\,,
\end{equation}
hence changing the frequency vector associated to the searched invariant tori.

The generic $r$-th normalisation step is performed in the same way provided that
the following non-resonance condition holds true:
\begin{equation}
\label{eq:res}
|\vet{k} \cdot \vet{\omega}^{(r-1)}| > 0\ , %
\qquad \forall\ \vet{k}\in\interi_2\setminus\{\vet{0}\}\ {\rm with} \ |\vet{k}|\le 2r\ .
\end{equation}
One can start from an expansion of the Hamiltonian $H^{(r-1)}$ of the
same form as in~\eqref{eq:hamexp}, where the upper index $0$ is
replaced by $r-1$. Hence, the generating functions $\chi_1^{(r)}$,
$\chi_2^{(r)}$ are introduced by solving the homological equations
obtained by replacing the upper index $1$ with $r$ in
formulas~\eqref{eq:kolgen1} and~\eqref{eq:kolgen2}.

The new Hamiltonian is therefore given by
\begin{equation}
H^{(r)}=\exp\Lscr_{\chi_2^{(r)}}\hat H^{(r)}\quad {\rm with}\quad \hat H^{(r)}=\exp\Lscr_{\chi_1^{(r)}} H^{(r-1)}\,.
\end{equation}

In order to better understand the ultimate goal of this algorithm
constructing invariant tori, let us suppose to be able to iterate it
{\it ad infinitum}. We would end up with a Hamiltonian of the form
\begin{equation}
\label{eq:haminfty}
H^{(\infty)}(\boldsymbol{p},\boldsymbol{q})\,=\, %
\boldsymbol{\omega}^{(\infty)}\cdot\boldsymbol{p} + \Oscr(\|\vet{p}\|^2)\ .
\end{equation}
Writing the equations of motion derived from the previous Hamiltonian,
it appears evident that the torus $\{\vet{p} =
\vet{0}\,,\ \vet{q}\in\toro^2\}$ is invariant.

\section{Parametric study on the $D_2$ parameter}
\label{sec:compalg}

By borrowing the techniques used in~\cite{Gio-Loc-San-2014} to ensure
the existence of elliptic tori for planetary systems, one could prove
the convergence of the algorithm described in the previous section
under very general conditions.  In practice, this means that: (i)~the
perturbation (ruled by~$\vet{I^*}$) is small enough; (ii)~the hessian
of the main quadratic term $f_2^{(0,0)}(\boldsymbol{p})$ is
non-degenerate; (iii)~the initial frequencies $\vet{\omega}^{(0)}$
belong to a suitable set having non-zero Lebesgue measure.

Here we do not investigate theoretically the convergence of the
algorithm that is instead numerically analysed. In the spirit of a
{\it reverse KAM approach}, we claim that some initial conditions
originate motions that are inside a stable region when the convergence
is evident from a numerical point of view.

We want to investigate the stability of extrasolar planetary systems
for the widest possible ranges of $D_2\,$ (\ie mutual inclinations)
and we want to take into account the uncertainties on other orbital
elements due to the observational limitations. Therefore, we have
found convenient to represent the coefficients of the expansions of
the Hamiltonians with intervals. Let us emphasise that such an
approach based on interval arithmetic allows us to cover completely a
set of values of the orbital elements.  This provides a key advantage
to the normal form approach with respect to the explorations purely
based on numerical integrations. In fact, when dealing with numeric
parametrical analysis the latter methods require to consider grids of
values of the initial conditions; moreover, the synthetic coverage
provided by the normal form approach (implemented with interval
arithmetic) is not possible.

When dealing with the proof of any KAM-type statement, it is essential to
establish an iterative scheme of estimates producing suitable
majorants. The ultimate goal of such a scheme is proving that the
norms of the two sequences of generating functions (\ie $\chi_1^{(r)}$
and $\chi_2^{(r)}$ in our settings) decrease exponentially. Therefore,
when such a behaviour is met in the plot of the norms of the generating
functions, this is the clear signature of the existence of invariant
KAM tori. In the case of a forced pendulum Hamiltonian model
(see~\cite{Cel-Gio-Loc-2000}), the study of the behaviour of
$\chi_2^{(r)}$ succeeded in extrapolating a good approximation of the
breakdown of the golden ratio invariant torus. As it has been
discussed in the Introduction, the existence of KAM tori implies the
long-time stability of the dynamics in a region surrounding them. Therefore,
this argument ensures that the stability in KAM sense is firmly
related to the convergence of the generating functions.

\begin{figure}[htpb!]
  \subfloat{\includegraphics[scale=.63]{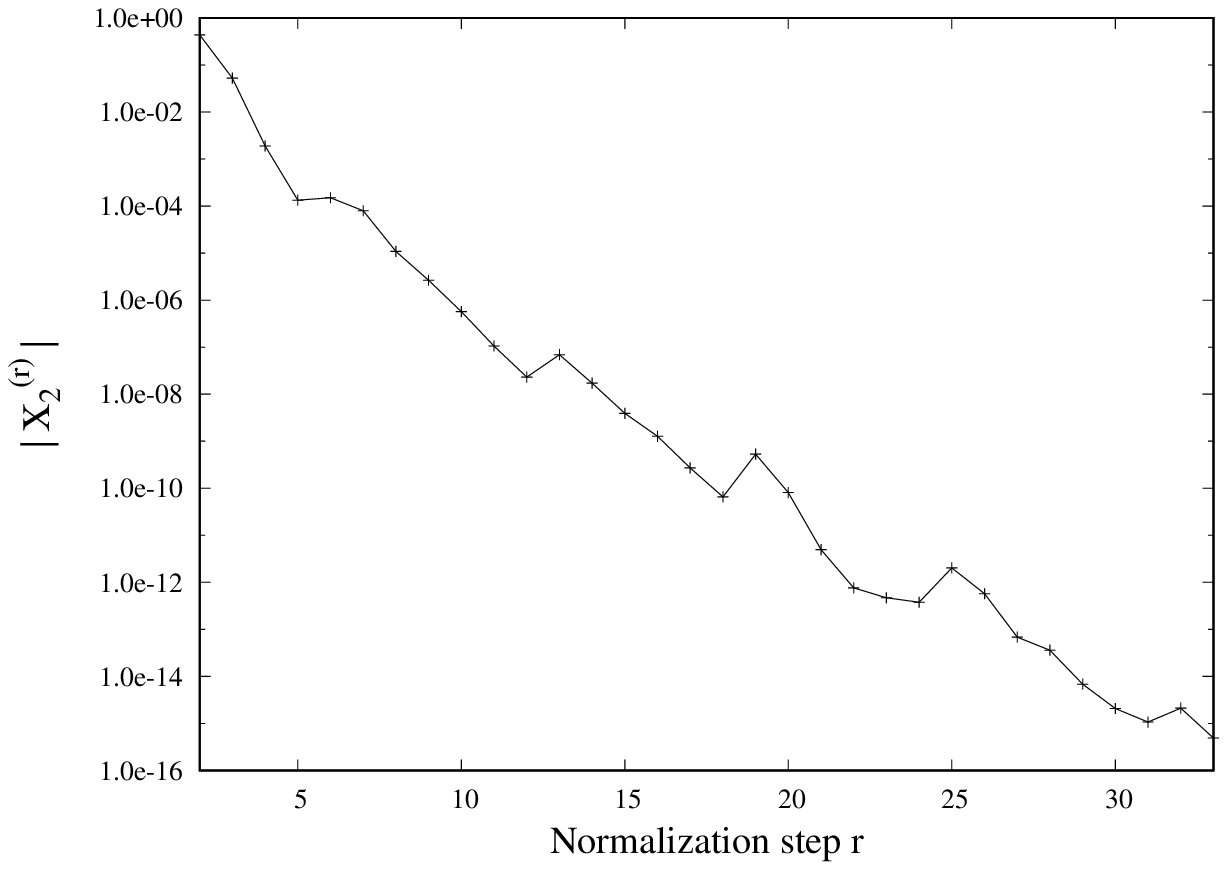}}
  \subfloat{\includegraphics[scale=.63]{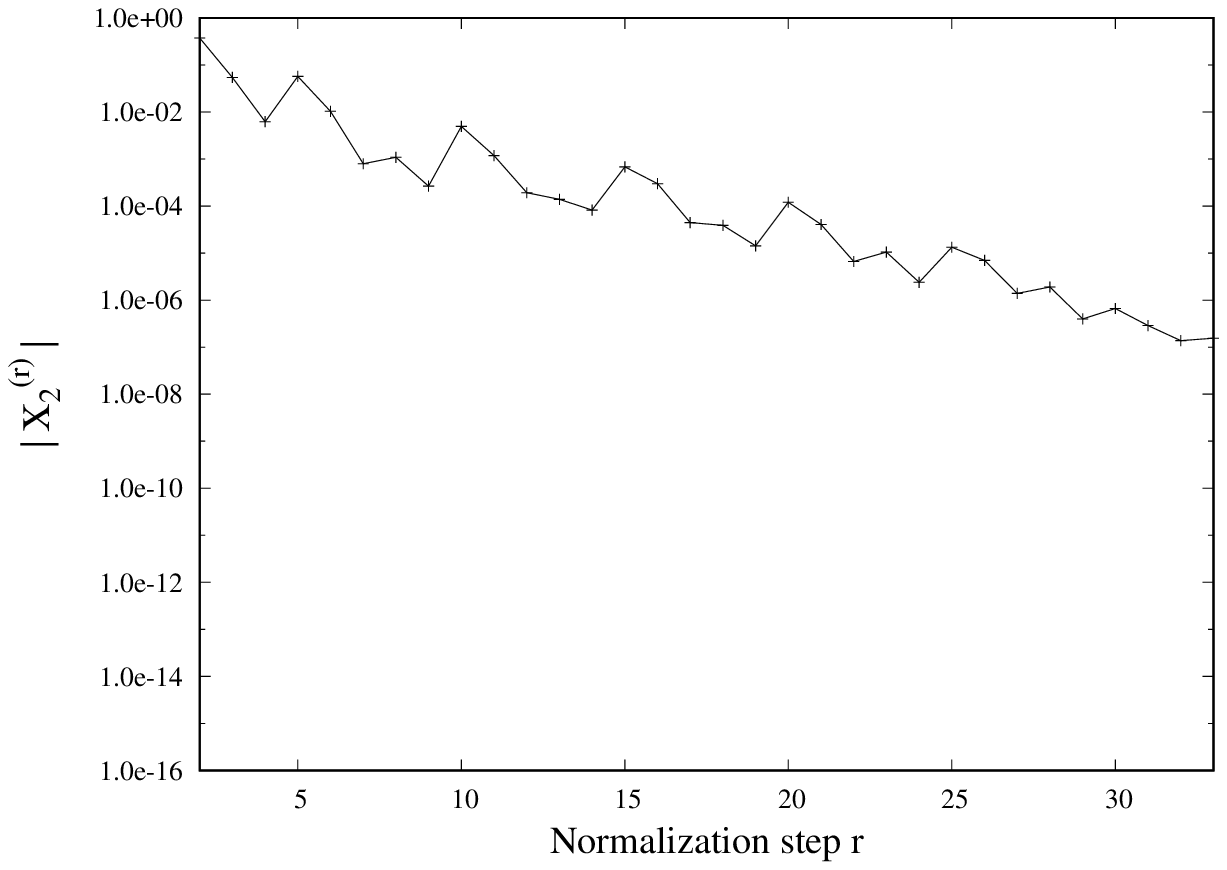}}
\caption{Results relative to HD 40307. Behaviour of the norms of the
  generating functions $\chi_2^{(r)}$ as a function of the
  normalisation step $r\,$. On the left, for values of $D_2\in
  [0.0164\,,\,0.0564]$. On the right for values of
  $D_2\in[0.0814\,,\,0.0864]$.  The orbital parameters of the system
  are listed in Tab.~\ref{tab:orbpar}.}
\label{fig:chi2}
\end{figure}

To fix the ideas, let us consider the specific case of the extrasolar
multiplanetary system HD~$40307$, whose orbital parameters are
reported in Tab.~\ref{tab:orbpar}. The plots of the norms of the
generating functions $\chi_2^{(r)}$ are shown in Fig.~\ref{fig:chi2}
for two different ranges of values of the parameter $D_2\,$. The norm
$\|\chi_2^{(r)}\|$ is nothing but the sum of the absolute values of
the coefficients of the terms appearing in its expansion. We decide to
focus on the behaviour of $\chi_2^{(r)}$ instead of the one of
$\chi_1^{(r)}$ because the former ones are usually bigger than the
latters. On the left hand side of Fig.~\ref{fig:chi2} we can
appreciate that the decrease of $\|\chi_2^{(r)}\|$ is sharp and quite
regular; we associate this behaviour to the convergence of the
algorithm. Often the algorithm crashes because the coefficients in the
expansions of the Hamiltonians inflate to the point where the
non-resonant condition~\eqref{eq:res} is not satisfied anymore. By
comparison, the decrease of the norms in the plot on the right of
Fig.~\ref{fig:chi2} is notably slower than the one on the left; for
instance, the norm of the last computed generating function on the
right is $6$ orders of magnitude bigger than the corresponding on the
left.

Obviously we aim to automatise the identification of the converging
procedures to avoid a visual inspection for each specific instance.
Having fixed the maximal normalisation order at $\bar{r}=33$, in our
codes the non convergence is established if at least one of the
following tests is true:
\begin{enumerate}
\label{enum:convtest}
\item the ratio $\|\chi_2^{(r)}\|\,/\,\|\chi_2^{(1)}\|$ is greater
  than $0.9^{\> r-1}$ for some $r$;
\item the norm $\|\chi_2^{(\bar{r})}\|$ is greater than $10^{-9}\>
  \|\chi_2^{(1)}\|$.
\end{enumerate}
Otherwise, we assume it is convergent.

\section{Results}
\label{sec:res}

In order to explicitly apply our approach, we selected extrasolar
systems where the eccentricities of the two major planets are small
(\ie less than $0.1$). In Tab.~\ref{tab:orbpar} we report the orbital
parameters of the systems considered: for the sake of simplicity in
the following we use as planetary masses the minimal ones listed
there.

\begin{table}[!]
\[
\begin{array}{cccccccc}
\toprule
\text{System} & {\rm Planet} & \thead{m\sin i\\ {[M_J]}} &
\thead{M_{Star}\\ {[M_{\odot}]}} & \thead{a\\ {\rm {[AU]}}} &
\thead{e\\ \phantom{+}} & \thead{\omega\\ {[^\circ]}} \\
\midrule
\text{HD}\>141399 & c & 1.33 & 1.14 & 0.704 & 0.048\pm 0.009 & 220\pm
40\\ & d & 1.18 & & 2.14 & 0.074\pm 0.025 & 220\pm
30\\ \text{HD}\>143761 & b & 1.045 & 0.99 & 0.228 & 0.037\pm 0.004 &
270.6\pm 6\\ &c & 0.079 & & 0.427 & 0.050 \pm 0.004 & 175\pm
125\\ \text{HD}\>40307\phantom{1} & c & 0.0202 & 0.77 & 0.081 &
0.060\pm 0.005 & 234\pm 1\\ & d & 0.0275 & & 0.134 & 0.070\pm 0.005 &
170\pm 10\\ \bottomrule
\end{array}
\]
\caption{Orbital parameters of the systems considered to apply the
  computational algorithm for the parametric study on $D_2$. For each
  column the unit of measure is reported in square brackets. The
  angle $i$ refers to the inclination of the orbital plane with respect
  to the line of sight.}
\label{tab:orbpar}
\end{table}

For the sake of completeness, we define some of the parameters ruling
the {\it finite size} of the expansions of the Hamiltonians introduced
in our formal algorithm (Secs.~\ref{sec:model}--\ref{sec:invtori}). In
Tab.~\ref{tab:parexp} we list the values of the integer parameters
$K_F$ and $N_S$ and of the mean motion resonance that is considered to
play the major role in the perturbation of the non-resonant fast
dynamics. Let us recall that $K_F$ gives the limitation on the
generating function $\chi_1^{(\Oscr 2)}$ that is needed to construct
the approximation of order $2$ in the masses; moreover, $N_S$ fixes
the maximal order in $e^2+i^2$ for the secular Hamiltonian $H^{({\rm
    sec})}$ (see Sec.~\ref{subsec:order2}). The series appearing
in~\eqref{eq:hamsecazang} and defining $H^{({\rm I})}$ is truncated at
the final value $s=15$; the same limitation is imposed on the
expansions of $H^{({\rm II})}$ and $H^{({\rm III})}$. Finally, the
maximal degree in the actions $\vet{p}$ is fixed at $4$ for the
expansions of all the Hamiltonians $H^{(r)}$ involved in the
normalisation up to order $\bar{r}=33$.

\begin{table}[!]
\[
\begin{array}{cccccccc}
\toprule
\text{System} & \text{Nearest Resonance} & K_F & N_S \\
\midrule
\text{HD}\>141399 & 5:1 & 12 & 8\\
\text{HD}\>143761 & 5:2 & 8 & 6\\
\text{HD}\>40307\phantom{1} & 2:1 & 6 & 8\\
\bottomrule
\end{array}
\]
\caption{Nearest resonance and values of the integer parameter
  $K_F$ and $N_S$ (as described in Section~\ref{subsec:order2}) for
  each system.}
\label{tab:parexp}
\end{table}

\begin{figure}[htpb!]
  \centering
  \subfloat{\includegraphics[scale=.63]{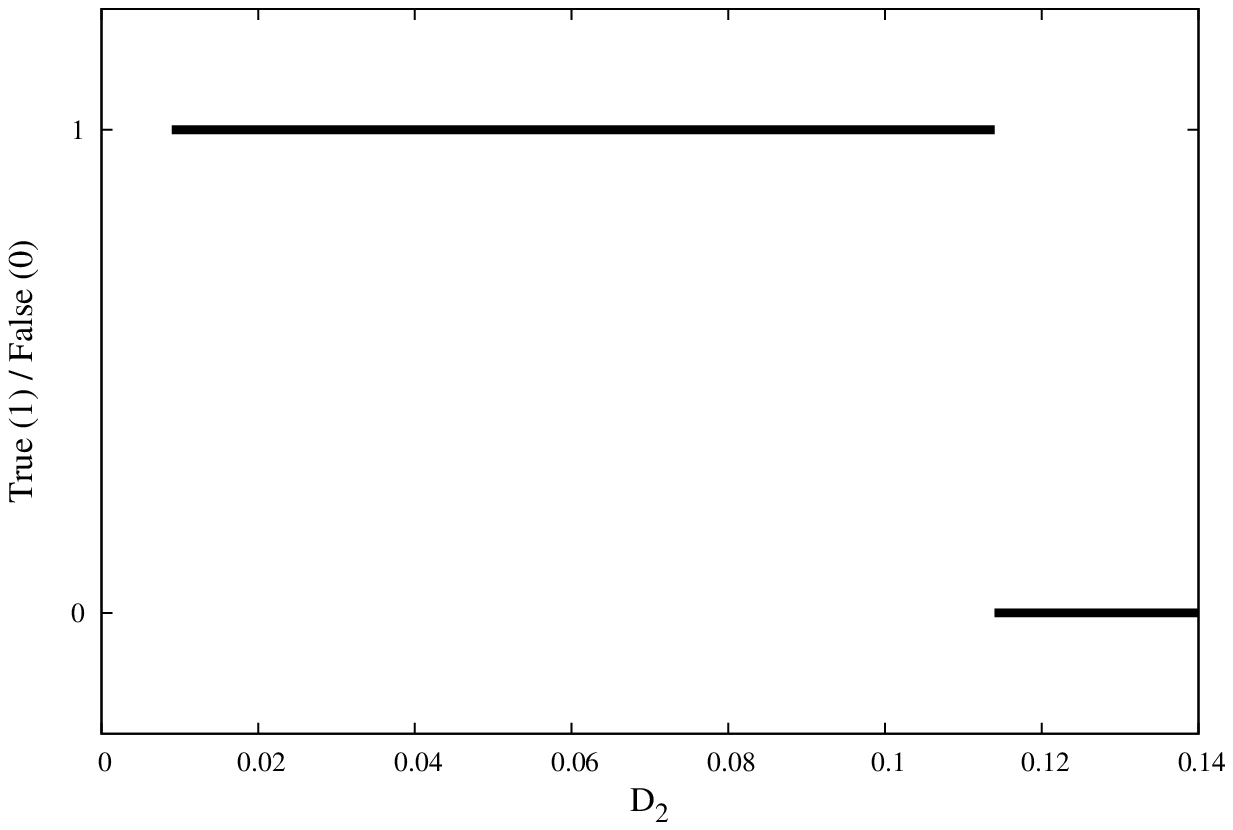}}
  \subfloat{\includegraphics[scale=.63]{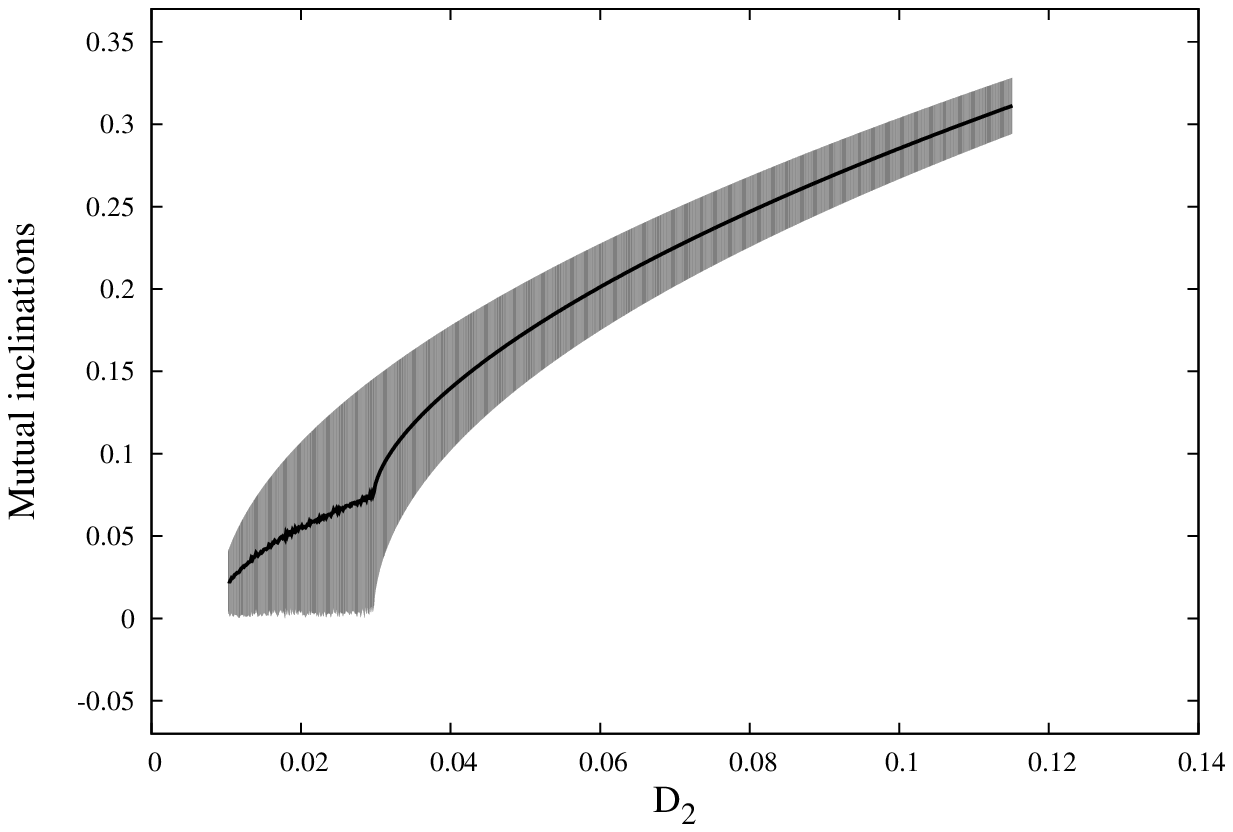}}
\caption{Results relative to HD~$141399$. On the left, the True/False
  output regarding the convergence of the algorithm. On the right, the
  range of values of the mutual inclination (in radians), where the
  thick line represent the mean value of the inclinations
  interval. Both the plots are drawn as functions of the parameter
  $D_2$.}
\label{fig:HD141399}
\end{figure}

\begin{figure}[htpb!]
  \centering
  \subfloat{\includegraphics[scale=.63]{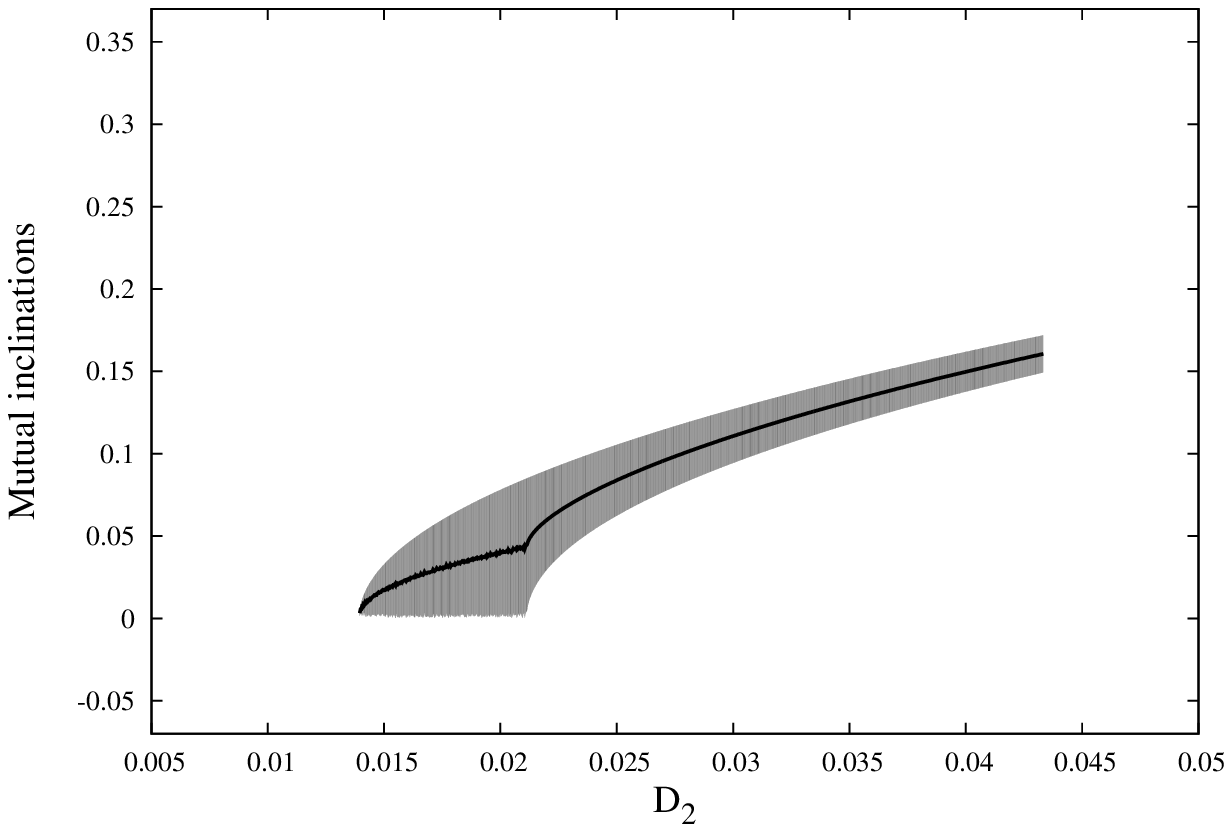}}
  \subfloat{\includegraphics[scale=.63]{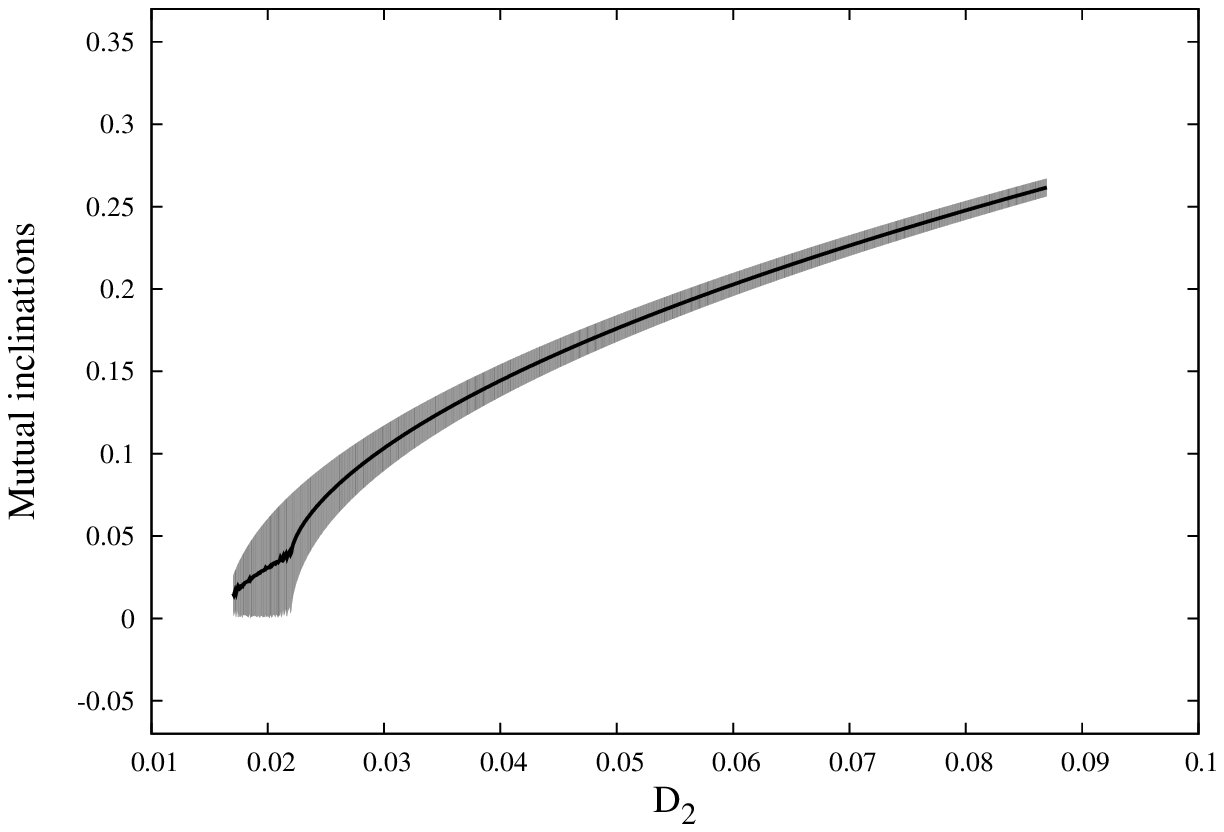}}
\caption{Plots of the mutual inclination as function of the
  parameter $D_2$. On the left, the results relative to
  HD~$143761$. On the right, those for HD~$40307$.}
\label{fig:imutHD}
\end{figure}

In Fig.~\ref{fig:HD141399} we present two plots relative to
HD~$141399$. On the left, we show the True/False output which results
from the tests on the convergence described in Sec.~\ref{sec:compalg}:
to each value of the parameter $D_2$ we assign $1$ if the system is
convergent, $0$ otherwise.  On the right, we show the plot of the the
mutual inclination as a function of the parameter $D_2$. By means of
the interval arithmetic, we can take into account the observational
errors on the orbital parameters of the system\eg the eccentricities
(as shown in Tab.~\ref{tab:orbpar}). Therefore, for each value of the
parameter $D_2$ we obtain a range of values for the mutual
inclination. For this reason, in all the plots concerning the mutual
inclination (right of Fig.~\ref{fig:HD141399} and
Fig.~\ref{fig:imutHD}) a central value with error-bars is drawn on the
$y$ coordinate. In Fig.~\ref{fig:imutHD}, we show the results for the
systems HD~$143761$ and HD~$40307$.

We can summarise the results provided by our implementation of the
Kolmogorov's normalisation scheme as follows: the systems HD~$141399$,
HD~$143761$ and HD~$40307$ are stable in the KAM sense, for mutual
inclinations up to $18^\circ$, $10^\circ$ and $15^\circ$,
respectively.  In this context, if we would have taken into account
the magnifying factor $1 / \sin i_j$ for the mass of the $j$-th
planet, we expect that the previous maximal mutual inclinations would
be slightly lower, except in the extreme case in which $i_j$ are close to
zero. Indeed, the main impact of considering larger masses would be
increasing the size of the correcting terms of order two in the masses
with respect to those of order one in the secular Hamiltonian
$H^{({\rm sec})}$.

\section{Conclusions and perspectives}
\label{sec:conc}

Up to our knowledge, this is the first application of KAM theory to
extrasolar planetary systems. As it is discussed in the previous
sections, actually we have not applied a statement of the KAM
theorem. Instead, we have exploited a keystone of the proof, \ie the
study of the convergence of the generating functions. In this respect
we can say that our approach is computer aided: the norms of some of
the initial generating functions are evaluated after having explicitly
calculated their expansions, instead of being analytically
estimated. The eventual convergent character of the constructing
algorithm in its entirety is inferred by the behaviour of said norms.
Our results should legitimately be included in the list of
the applications of KAM theory to realistic physical models
(see\eg~\cite{Celletti-1994, Cel-Chi-2007, Gab-Jor-Loc-2005}). In
fact, for what concerns the tori that are invariant with respect to
the secular Hamiltonian and characterized by the complete circulation
of the arguments of the pericenters, the values of the mutual
inclinations for which the Lidov-Kozai resonant region takes place can
be considered as a natural upper limit\footnote{In the Laplace plane
  frame, the region of the Lidov-Kozai resonance is characterized by
  the libration of the argument of the pericenter of the inner planet
  (see~\cite{Lidov-1962}).  The implicit adoption of such a frame has
  been essential in order to perform the reduction of the angular
  momentum sketched in Sec.~\ref{subsec:exp}.  Therefore, the
  comparison between our results and those for that resonant region is
  valid because also our Hamiltonian model is written in the secular
  canonical coordinates with respect to the Laplace plane.}. In
extrasolar systems such a critical value of the mutual inclinations is
usually located at about $40^\circ$
(see~\cite{Lib-Tsi-2009}). Therefore, for the three systems here
considered, our results about the stability in the KAM sense cover a
set of values whose extension ranges between $25\%$ and $50\%$ of the
maximal one.

We shall now point out the weaknesses of our approach. Our
constructing algorithm does not work when the eccentricities of the
planets are not small. In fact, the procedure has generated divergent
series when it has been applied to the systems HD~$109271$,
HD~$155358$ and HD~$4732$; in all of them there is at least one of the
planets whose eccentricity is between $0.1$ and $0.25$. Thus, it seems
that our approach is limited to systems with planetary eccentricities
$< 0.1$. Since we are able to produce results for small inclinations
of the major planets of the systems, the ideal situation is very
similar to that of our Solar System. This is not surprising, since the
whole approach has been adapted from the one described
in~\cite{Loc-Gio-2000}, which in turn has been tailored to the Jovian
planets. In particular, the series expansion of the three-body
planetary Hamiltonian is in power series of some coordinates and
parameters that are of the same order of the eccentricities and the
inclinations.

A natural goal for the future would be to remove the limitations
affecting the approach described in this paper. We think that some of
them are intrinsic in the definition of stability that we assumed.
Actually, since the beginning we postulated that the motions of the
major planets are quasi-periodic and their orbits lie on KAM tori
constructed with expansions in small eccentricities and
inclinations. Such a prescription is extremely strict. In our opinion,
any substantial improvement of the method will be based on a clever
weakening of the requirements. This should be done by identifying a
suitable integrable approximation of the secular dynamics that can
be shown to be convergent even for large eccentricities. In the very
different context of the orbits of the Trojan bodies, this change of
attitude has been shown to produce substantial enhancements
(see~\cite{Paez-Loc-2015, Paez-Loc-Eft-2016}). In future
works, we plan to extend this kind of ideas to the problem of
determining values of the inclinations consistent with (a suitable
type of) stability.

\subsection*{Acknowledgements}
During the early stages of preparation of this work, U.L. has been
financially supported by the project ``DEXTEROUS - Uncovering
Excellence 2014'' of the University of Rome ``Tor Vergata''.
\paragraph{}{\it Every author has no conflicts of interest to declare.}

\end{document}